\documentclass[conference]{IEEEtran}
\IEEEoverridecommandlockouts
\usepackage{cite}
\usepackage{amsmath,amssymb,amsfonts}
\usepackage{graphicx}
\usepackage{textcomp}
\usepackage{xcolor}
\def\BibTeX{{\rm B\kern-.05em{\sc i\kern-.025em b}\kern-.08em
    T\kern-.1667em\lower.7ex\hbox{E}\kern-.125emX}}

\usepackage{pgfplots}
\usepackage{url}
\usepackage{hyperref}
\usepackage{algorithm} 
\usepackage{algpseudocode} 
\usepackage{enumitem} 
\usepackage{xspace} 
\usepackage{float} 
\usepackage{makecell}

\pgfplotsset{compat=1.18} 

\begin{document}

\title{Vector embedding of multi-modal texts: a tool for discovery?\\
%{\footnotesize \textsuperscript{*}Note: Sub-titles are not captured in Xplore and should not be used}
\thanks{This research funded in part through a grant from the National Science Foundation OAC-2209872. This research reflects the views of the authors and not of the NSF.}
}

\author{\IEEEauthorblockN{Beth Plale}
\IEEEauthorblockA{\textit{School of Computer and Data Sciences} \\
\textit{University of Oregon}\\
Eugene, OR USA\\
0000-0003-2164-8132}
\and
\IEEEauthorblockN{Sai Navya Jyesta}
\IEEEauthorblockA{\textit{Computer Science Engineering Dept}\\
\textit{Indiana University}\\
Bloomington, IN USA\\
sjyesta@iu.edu}
\and
\IEEEauthorblockN{Sachith Withana}
\IEEEauthorblockA{\textit{Intelligent Systems Engineering Dept} \\
\textit{Indiana University}\\
Bloomington, IN USA\\
swithana@iu.edu}
}

\maketitle

\begin{abstract} 
Computer science texts are particularly rich in both narrative content and illustrative charts, algorithms, images, annotated diagrams, etc. This study explores the extent to which vector-based multimodal retrieval, powered by vision-language models (VLMs), can improve discovery across multi-modal (text and images) content. Using over 3,600 digitized textbook pages largely from computer science textbooks and a Vision Language Model (VLM), we generate multi-vector representations capturing both textual and visual semantics. These embeddings are stored in a vector database. We issue a benchmark of 75 natural language queries and compare retrieval performance to ground truth and across four similarity (distance) measures. The study is intended to expose both the strengths and weakenesses of such an approach. We find that cosine similarity most effectively retrieves semantically and visually relevant pages. We further discuss the practicality of using a vector database and multi-modal embedding for operational information retrieval. Our paper is intended to offer design insights for discovery over digital libraries.
\end{abstract}


\section{Introduction}

Large-scale digital repositories, such as the HathiTrust with its 17 million plus digitized volumes from research libraries around the world, can be valuable resources for scientific discovery.  With digitized collections running into the 10's of millions of volumes (books, periodicals, monographs, etc.), material dating back hundreds of years, and issues of copyright that limit access rights, finding the right material for a particular research or educational use can be a challenge.

Document retrieval is a foundational need as materials discovery is the first step in any subsequent analysis of digital materials.  The foundational nature is reinforced by the FAIR principles (Findable, Accessible, Interoperable, and Usable)~\cite{FAIR} where the principle of Findability is assumed by the other principles (accessible, interoperable, etc.) Discovery is most commonly, at least until recently, been supported through full text search indexes. Implemented as an inverted index, full text search indexes are fast and scale well in support of full-text search. Access is through a query language such as Lucene (for the Solr search index).  

With the recent release of Large Language Models (LLM), the LLM has quickly become a useful discovery vector by being able to act remarkably well on human language stated requests for content. The use of LLMs for discovery of content that is under copyright is a matter of considerable societal interest at the time of this writing.   Karamolegkou et al. (2023) ~\cite{copyrightLLM} find that "larger language
models memorize at least a substantial repository of copyrighted text fragments." A recent Quarles law firm newsletter discusses two recent district court rulings by the Northern District of California on the sources of the data (pirated or gained through legitimate agreements) and on fairness of use. The rulings provide evidence that leans towards LLM training as transformative and thus accommodated under fair use. 

Our efforts are more narrowly defined. We focus on one component of the LLM ecosystem, the vector database, and study whether that component can augment a digital library discovery strategy. Specifically, can numerical representations of data and images and distance measures between these numerical representations reveal relationships that are otherwise non-obvious? 
%The context of our study is large digital libraries of digitized volumes from research libraries.  

Our approach is to vectorize digital content (represented as numerical vectors) which we postulate removes concerns about reverse engineering and thus violate copyright restrictions.  Instead of vectorizing the OCR'ed pages, which have no graphics, we vectorize page images.
%and utilize LLMs' natural language interactions to aid discovery.  
Our goal is to assess how well the vector database is at finding pages that our manual efforts at constructing ground truths has revealed.  And to assess the serendipity factor, that is, where the vector database returns pages that are outside our ground truth.  %pre-determined ground truth evaluate where this approach adds to the overall discoverability landscape for large-scale digital libraries. That is, 
Overall we aim to contribute new knowledge for others as they assess the viability of using vector embeddings of page images for purposes of discovery. 
%assessment one may make of the  the contributions that could be made to discovery over large scale digital libraries through vector embeddings of multi-modal digitized texts. 
Our ultimate goal is to assess the extent to which a vector database will add to the discoverability of a digital library beyond retrieval techniques currently in place. This study is a step along that path. 

Other approaches to making digital content accessible to discovery and other needs is the HathiTrust Research Center Extracted Features (EF) Dataset~\cite{9651860}. While not exclusively for discovery, the EF dataset consists of structured data and metadata drawn from OCR'ed volumes of the HathiTrust Digital Library.  The dataset has enabled unique computational analyses of the corpus and is released under a CC by 4.0 license.

%LLMs offer improvements to query language solutions. Through natural language queries, LLMs promise to more accurately reflecting human intent behind the discovery questions being asked.  We seek to exploit natural language queries to enhance discovery over content found in digitized college level computer science textbooks.  

Our study uses technology textbooks in distributed systems, database, computer networks, AI and operating systems because these texts make extensive use of diagrammatic graphics flowcharts, hierarchical graphs, process flows, UML diagrams, and annotated graphs.  Our dataset is just over 3600 pages (pdfs) from digitized technology textbooks, specifically, textbooks in distributed systems, computer networks, AI, programming languages, and operating systems. Also included in the collection are 116 digitized recipes from a book of recipes, descriptions, and pictures. Every page is rendered as an image with a DPI of 300, considered a high resolution image standard.

We ask, for instance, a natural language query for a distributed systems concept (such as "concurrency"). Is the result satisfied by the body of the narrative text or can also have its result set enhanced by information available in related diagrammatic graphics? When querying for a concept that is described in the narrative of a document, how good is the system at also return relevant diagrams? These are multi-modal queries where the expected result set draws from both text and graphs.

What we aim to discern from this study is whether the information in diagrammatic graphics can be used together with text to enhance discovery through natural language queries. Making sense of information in multiple forms has been explored under the banner of multimodal machine learning, a field which has been active since the late 2010's with some of the early work by Collell et al. ~\cite{Collell_Zhang_Moens_2017}, and which we draw on for this study.

We study how to effectively utilize available multimodal cues from diagrams for the cross-modal diagram-text retrieval task.  Specifically, we apply tools for multimodal embedding, where both text and images are represented via the same mathematical object (that is, a vector). 
Multimodal embeddings represent multiple data modalities (e.g., text, graphics) in the same vector space such that similar concepts are located in vector space proximity to one another. 
This means that multimodal data resides in a single vector database.

Each page image is divided into 1030 visual patches and features extracted about each patch.  The extracted features are then projected into a shared 128-dimensional embedding space. The tool to create the patches and project the patches into 128-dimensional space is the ColPali\cite{faysse2025colpaliICLR} model. ColPali is a Vision Language Model to produce multi-vector embeddings from images of document pages. ColPali uses the PaliGemma-3B model\cite{2024paligemma}, which combines SigLIP\cite{alabdulmohsin2023getting} patch embeddings with a Gemma-2B language model\cite{gemma2024}, to generate ColBERT-style\cite{khattab2020colbert} multi-vector representations of text and images. 

We evaluate the system’s effectiveness at retrieval tasks by measuring its ability to retrieve relevant documents for a query. We set a threshold of the top-k best results, which we set at $k=5$ and call \textit{Top-5}.  For each query in the benchmark that we have developed we ask the system for the Top-5 ranked results based on ground truth and compute the chosen metric(s) over that set. The  \textit{ground truth} is a triplet of query, passage, and relevance judgment; it is per query and is curated manually. That is, every query has a predetermined ground truth (sometimes called "gold standard") result.  

The contribution of this paper is the study. We design a benchmark set of queries, and carefully curate ground truth results for each. We fin that queries over text and image features make it easier to uncover relevant material, especially where diagrams and visuals are essential for understanding. Our study also shows that cosine similarity works better than other distance measures for keeping the meaning of both text and images consistent, which leads to more reliable retrieval results. We are interested in results that are unexpected. While we carry out an early experiment (experiment 4), more work is needed.  Taken together, these insights show practical ways to strengthen systems like HathiTrust, particularly for collections that mix text with visually rich content, and highlight how multimodal retrieval can play a key role in future improvements.

%First, Cosine Similarity provides the most consistent semantic alignment across both factual and abstract queries, despite Dot Product scoring slightly higher on raw metrics. Second, the Semantic-10 benchmark reveals retrieval challenges that are not captured in standard fact-based queries, validating its role as a necessary complement to Baseline-75; and Third, Cosine Similarity frequently retrieves relevant but unlabeled content, demonstrating that rigid ground truth labels may underrepresent the true effectiveness of semantic vector retrieval, particularly in multi-page document collections. 

%These findings underscore the importance of multimodal benchmarks and semantically aware similarity functions for next-generation document discovery systems, particularly at the scale of digital libraries like HathiTrust. 

%Our study narrowly focuses on technology texts because of their rich use of charts and other graphics. However, we discuss factors such as scalability of the solution to far larger collections and additionally ask questions such as the impact of ever-expanding context windows of LLMs on the relevance of our findings. 

The remainder of the paper begins with use cases in Sec~\ref{sec:usecase}.  The methodology of the study follows in Sec~\ref{sec:method} and the experimental evaluation in Sec~\ref{sec:eval}. We then discuss the findings in Sec~\ref{sec:discussion}.  The paper concludes with Related Work, Sec ~\ref{sec:related} and  open questions and links to associated benchmarks and code in Sec~\ref{sec:conclusion}.

\section{Use Cases}\label{sec:usecase}
%Modern digital libraries like HathiTrust preserve and provide access to millions of digitized volumes across the wealth of human knowledge as held in research libraries. 

 We illustrate our study through three representative queries from our benchmark one each for multi-page, multi-modal, and concept-driven information needs. 

For each of the queries, we explain the query and then discuss the top 5 results that are retrieved under cosine similarity.  The purpose is to illustrate the kinds of similarities that are uncovered through vector embedding.
%In each case, we show that relevant material was retrieved by the system but was not included in the original gold standard, highlighting the limits of static evaluation and the value of flexible semantic retrieval in a library context.

\begin{itemize}
    \item 
\textit{“Tell me about the Bellman-Ford algorithm and give me an example”}  \\
One would expect here a description of how the algorithm works, possibly a pseudocode representation and either a pseudocode example or narrative form example. As pseudo-code algorithms can take up considerable space on a page, it is reasonable to expect that the results are presented across multiple consecutive pages. This query, MP-14, is thus categorized as a multi-page query in our benchmark.

Of the top 5 pages returned, the Bellman-Ford algorithm's fundamental theoretical framework is presented in the first page that was retrieved, outlining important ideas such as state space, goal states, actions, and transition models. A detailed and well-worked example demonstrating the algorithm's practical application is provided on the next page. 
%The recovered content is extremely relevant due to the conceptual connection between the application and the theoretical underpinning. This outcome offers a comprehensive understanding of the algorithm's operation by demonstrating the system's capacity to manage multi-page searches and obtain both the essential practical example and the underlying explanation.

\item
\textit{“How do state transition graphs work?”} \\ 
State transition graphs show the logical progression of an computational entity through states.  The query, MM-1, is categorized as a multi-modal query as the expected result is an image and description. While the gold standard included one page with a labeled diagram, our system retrieved a second page containing a paragraph-length textual explanation of each state and its transition. The conceptual alignment between the text and the query intent made the second page highly relevant. This result illustrates the system’s ability to connect visual and textual signals, which is particularly important in illustrated technical textbooks.

\item{\textit{“How does eventual consistency differ from strong consistency in terms of behavior and guarantees?”}} \\
This query, CP-13, is an example of a concept-driven query.  Eventual consistency is a distributed systems concept that has to do with computer processes that are geospatially distributed but have the requirement to have a consistent view of some piece of information.  Eventual consistency allows some time before the world views become consistent.

This query aims to highlight the differences between eventual consistency and strong consistency, both in terms of behavior and guarantees.  The first page retrieved provides a fundamental overview of eventual consistency, detailing its key ideas and use in distributed systems.  The second page retrieved provides a more detailed explanation of strong consistency, including its tougher assurances and the behavior it ensures in data synchronization between systems.  The theoretical foundations of both consistency models are aligned; therefore, the retrieved content is extremely relevant to the query.  This result demonstrates the system's capacity to handle concept-driven inquiries by obtaining content that both explains and contrasts each consistency model, providing a thorough grasp of their distinctions and implications in distributed system design.
\end{itemize}

\section{Methodology}\label{sec:method}

\begin{figure}[htbp]    % Changed to figure* for full width
    \centering
    \includegraphics[width=7cm]{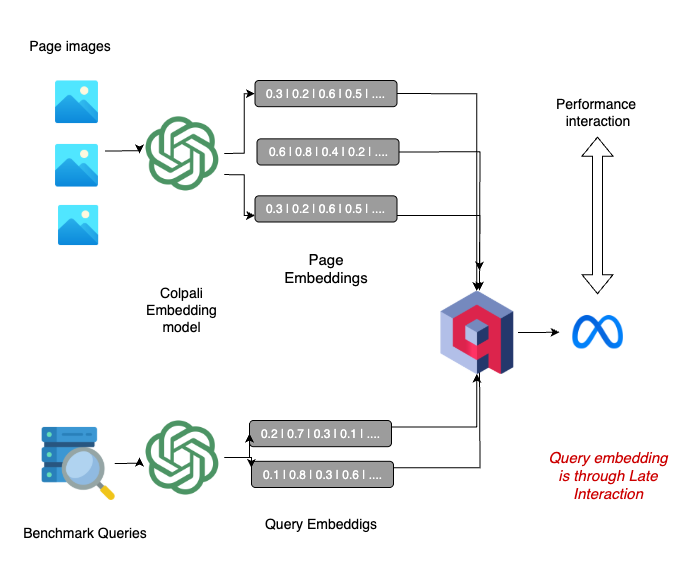}
    \caption{Architectural Schematic}
    \label{fig:workflowDiagram}
\end{figure}

%To evaluate the effectiveness of various similarity functions in the context of semantic document retrieval, 
We design a modular and extensible benchmarking framework centered around the ColPali multimodal embedding model. 
The framework, shown in Figure \ref{fig:workflowDiagram}, has page embeddings across the top. This happens first.  Using the ColPali multimodal embedding model, vector embeddings are created for each page in the 3600 page dataset. Adoption of ColPali eliminates the need for complex preprocessing pipelines involving OCR, layout detection, and chunking. Each page is processed through a vision-language embedding model (PaliGemma-3B) and resulting embeddings are stored in Qdrant, a vector database supporting multi-vector indexing and ColBERT-style late interaction search. 

Along the bottom of the figure, and occurring at a later time, each query is passed through ColPali to create its vector embedding which is then stored to Qdrant. From there, each query is asked through an LLM which consults the Qdrant database and returns a pre-specified top number of relevant pages that have a minimal distance from the query using a pre-defined distance measure. The top relevant pages are selected based on similarity scores between the query and document embeddings, computed using one of four distance functions (Cosine, Dot Product, Euclidean, Manhattan). 
 Each query produces a ranked list of pages, from which the top-k results are returned for evaluation. 

The design is guided by the need for rigorous experimental control, where all aspects of the retrieval pipeline are held constant across all experiments, including embedding model, dataset, indexing method, query formulation, and evaluation protocol. The only variable that can be altered between runs is the choice of similarity function used during search of the vector database. This strict isolation allowed us to attribute observed performance variations solely to the behavior of the distance metric in high-dimensional semantic space.  This is an important factor as one of our evaluations is to systematically assess how different vector distance metrics (i.e., Cosine similarity, Dot Product, Euclidean distance, and Manhattan distance) affect the quality and relevance of retrieved results.

Queries are issued to each configuration independently, and retrieved results compared against a manually curated ground truth using standard information retrieval metrics. 

\subsection{Pipeline} 

The pipeline is made up of three sections: document indexing, image retrieval, and generation.  We discuss each in turn. 

\paragraph{Document Indexing} We bring a dataset of 3,612 image-based textbook pages and a small recipe book that is rich with pictures. The textbooks are five popular undergraduate textbooks in computer science where we have rights to the digital copy. The text books are in operating systems, programming languages, computer systems, and AI. 

%\begin{itemize}   \item %\textit{Operating System Concepts} (10th Edition) –     operating systems concepts and images of process management, memory allocation, and file systems.
% \item %\textit{Artificial Intelligence: A Modern Approach} (4th Edition) – state-of-the-art 
 %   AI, machine learning, planning, and search abstractions, processes.
  %  \item %\textit{Introduction to Automata Theory, Languages, and Computation} –  formal languages, automata, grammar, and Turing machines.
%  \item %\textit{Principles of Computer System Design} –    layered architecture, abstraction, system interfaces, and fault tolerance.
%   \item %\textit{Cook’s Illustrated Annual 2019 Recipes} –  food recipes. Included as a contrasting domain to evaluate model generalization across non-technical, visually structured content.
%\end{itemize}

Each volume is originally in the form of a single document in PDF format. 
%and was preprocessed through the following steps to create a consistent input format for our retrieval system:
%Each page (PDF) is split into individual pages using \texttt{PyMuPDF}, ensuring that page boundaries and formatting were preserved.
Using \texttt{PyMuPDF}, the full volume PDF document is broken into individual pages where each page is rendered as an image with a DPI of 300, which is considered a high resolution image standard.  DPI of 300 balances readability of textual and visual elements with computational efficiency. The resulting image is stored as a PNG image; PNG ensures lossless compression and layout fidelity, which is essential for downstream visual embedding.

The page images are fed to Colpali, where they are processed by the \textit{PaliGemma-3B} model, which is part of Colpali. PaliGemma-3B (specifically PaliGemma-3Ba) 
%state-of-the-art multimodal vision-language embedding system. PaliGemma-3B integrates SigLIP for patch-level visual representation and Gemma-2B as the text reasoning backbone. The model 
divides each image into 1030 visual patches, extracts features from each patch, and projects them into a shared 128-dimensional embedding space. This results in a fixed-length matrix representation of size $1030 \times 128$ for each page, capturing information about the textual and structural aspects of the page as well as about the diagrams.

All page embeddings are stored (indexed) to a vector database. We selected the \textit{Qdrant Vector Database}~\cite{qdrant2024} as it is high-performance, open source, and allows for multi-vector indexing which means each document (\textit{i.e.}, page) can be stored as a sequence of vectors, rather than a single aggregated representation. 

The key to our scalable approach is a technique pioneered in 2020 by the ColBERT information retrieval model ~\cite{khattab2020colbertefficienteffectivepassage} called \textit{Late Interaction Scoring (LI)}. LI is a retrieval architectural design in which query and document representations are processed independently and only interact at the final stages (when scoring is applied to identify the most relevant pages). Unlike early interaction models that fuse query and document embeddings before or during encoding (e.g., through cross-attention layers), LI models maintain separate encoding pipelines, enabling efficient storage and reuse of document embeddings. The core idea is to represent both queries and documents as sequences of vectors (e.g., token embeddings for text or patch embeddings for images), and compute fine-grained relevance scores through token-to-token or token-to-patch similarity comparisons.

Specifically, in ColBERT, and later ColPali, late interaction works as follows: a document is scored by taking the maximum similarity between each query vector and all document vectors, and then aggregating these maxima into a single relevance score. Applied to our system, each textbook page is represented by a fixed set of visual patch embeddings, while each natural language query is transformed into a sequence of dense token embeddings. At retrieval time, each query token interacts with all document patches via a configurable similarity metric (e.g., cosine, dot product), and a late aggregation operation (e.g., max-sum pooling) produces the final document relevance score.

The late interaction approach enables token-level alignment between query concepts and fine-grained visual features of the page (e.g., specific diagrams, formulas, or text blocks), making it well-suited for visually dense documents like textbooks. Moreover, late interaction scoring allows us to isolate and evaluate the effect of different similarity functions (cosine, dot product, Euclidean, Manhattan) within a controlled and interpretable retrieval pipeline.

%Late Interaction Scoring enables fine-grained token-level interactions during retrieval, essential for ColBERT-style late interaction scoring.

%The preprocessing pipeline ensured uniformity in input format and semantic representation across the dataset, enabling controlled experimentation with retrieval functions and accurate comparison across similarity metrics.

\paragraph{Image Retrieval}   
%Each metric is used to perform independent top-$k$ searches over the same dataset and embeddings. 

To evaluate the impact of different similarity measures on retrieval performance, we configure four distinct vector distance functions within the vector database. 
Each distance function is configured for use in a separate Qdrant collection having identical content and configuration to the others except for the distance function. All other components—including embeddings, the $Top-5$ value, retrieval logic, and evaluation metrics— are held constant to ensure that the effect of the similarity function can be analyzed in isolation.

\begin{itemize}

    \item \textbf{Cosine Similarity:}  
    Cosine similarity measures the cosine of the angle between two vectors. It is invariant to vector magnitude, making it well-suited for high-dimensional semantic representations such as text or image embeddings. It emphasizes direction over magnitude.

    \[
    \text{CosineSim}(\mathbf{A}, \mathbf{B}) = \frac{\mathbf{A} \cdot \mathbf{B}}{\|\mathbf{A}\| \|\mathbf{B}\|}
    \]

    \item \textbf{Dot Product:}  
    Dot Product computes the raw inner product of two vectors. Unlike cosine similarity, it is sensitive to the magnitude of vectors. It can be useful when the scale of embeddings carries semantic meaning, though it may bias results toward longer vectors.

    \[
    \text{DotProduct}(\mathbf{A}, \mathbf{B}) = \sum_{i=1}^{n} A_i \cdot B_i
    \]

    \item \textbf{Euclidean Distance (L2):}  
    Euclidean distance calculates the straight-line distance between two points in vector space. It penalizes both magnitude and direction differences. It is effective in low-dimensional or dense feature spaces but may be less robust in high-dimensional sparse embeddings.

    \[
    d_{\text{Euclidean}}(\mathbf{A}, \mathbf{B}) = \sqrt{\sum_{i=1}^{n} (A_i - B_i)^2}
    \]

    \item \textbf{Manhattan Distance (L1):}  
    Manhattan distance, also known as city-block distance, sums the absolute differences between vector components. It is less sensitive to outliers than Euclidean distance and can be more appropriate for sparse vectors.

    \[
    d_{\text{Manhattan}}(\mathbf{A}, \mathbf{B}) = \sum_{i=1}^{n} |A_i - B_i|
    \]

\end{itemize}

\paragraph{Generation}

With the page embeddings indexed in the vector database, queries from the benchmark are issued against the database. 
After applying the similarity function, the system scores the retrieved pages based on how closely they match the query. This step helps identify the most relevant pages from the initial set of results. The system ranks the pages by comparing the query with each one and determining their relevance.  For each query, returned are the top relevant pages ($Top-k$).  We configure the database to return 25 results, then use a local algorithm to identify the top 5. The local algorithm is simple; it merely removes duplicates where they occur.

Max-sum pooling is used to look at each part of the page and select the part that matches the query the best. The best matches are combined to get a final score for the page. This ensures the system focuses on the most relevant parts of each page.

Once the initial scoring is done, the next step is re-ranking, which fine-tunes the results. Here the order of documents is adjusted based on how highly relevant they are. This helps make sure that the pages that are most likely to be useful to the user are shown first.

\section{Evaluation}\label{sec:eval}

We carry out four experiments. The first evaluates the different similarity measures using the full 75 query benchmark. The second experiment works with just a subset of the benchmark, that is, the 14 multi-page queries. This is intended to assess distance measure performance on a single homogeneous query category.  In the third experiment we use a different subset of queries, this being the conceptual query subset and again examine distance measure results for this homogeneous category.  The fourth experiment uses the full benchmark to study results that fall outside the Top-5 results.  These latter results we conjecture could hint at new serendipitous relationships.   

Our experimental environment is a modular set of services: 1) Model Server is a standalone embedding service using PaliGemma-3B. 2) Qdrant Database contains four vector collections with identical content and isolated distance metrics. 3) Python code controls query orchestration, vector search, and metric logging, and 4) Frontend UI which is a Streamlit interface for natural language queries and real-time result visualization.  

We isolate the choice of distance function to use by fixing other variable aspects: the embedding model, index structure, and query formulation. This ensures that the observed performance differences are attributable solely to the choice of distance function. 

All experiments are run on an Intel i7-7800X CPU, 32GB RAM, NVIDIA Quadro P1000 GPU.

\subsection{Query Benchmark} We design a benchmark made up of 75 natural language queries to cover a range of retrieval challenges. These are organized into categories that span from simple text lookups to complex multimodal reasoning tasks, see Table~\ref{tab:query-benchmark2}.

%when visualizing results, take 1 representative query from each category.  Set "top" to infinity. plot with distance on Y axis and data points for each response.  There will be 5 tables as result.  Drop these into paper and give the query that you used, and \# responses. 

\begin{table}[htbp]
    \caption{Query benchmark categories and counts}
       \label{tab:query-benchmark2}
    \centering
    \renewcommand{\arraystretch}{1.25}
    \setlength{\tabcolsep}{2pt}
    \resizebox{0.45\textwidth}{!}{
        \begin{tabular}{|l|p{6.3cm}|c|}
            \hline
            \textbf{Query Type} & \textbf{Description} & \textbf{Count} \\
            \hline
            Visual & Refers to diagrams, flowcharts, or illustrations conveying conceptual content. & 9 \\
            Textual & Asks for definitions, explanations, or semantic details in natural language. & 14 \\
            Multi-Modal & Requires understanding of both visual and textual content, such as labeled figures with descriptions. & 10 \\
            Tabular & Involves structured data such as parameter tables, configuration matrices, or protocol listings. & 7 \\
            Numerical & Focuses on retrieving numeric values like thresholds, performance metrics, or quantitative comparisons. & 8 \\
            Multi-Page & Requires retrieval of conceptually linked content spread across multiple pages (e.g., algorithm + figure). & 14 \\
            Conceptual & Focuses on abstraction and higher-level reasoning, requiring the model to link visual and textual domains for understanding complex concepts. & 13 \\
            \hline
            \textbf{Total} & & \textbf{75} \\
            \hline
        \end{tabular}
    }
\end{table}

\textit{Visual queries} are intended to reference  diagrams, flowcharts, or images that convey conceptual content, such as “Show a figure of process synchronization and its different states.” These queries test the model’s ability to retrieve visual layouts embedded within pages. \\
\textit{Textual queries} are such as “What conditions lead to a deadlock” assess semantic understanding using traditional natural language search. \\
\textit{Multi-modal queries} require retrieval of both visual and textual content, such as “How does the reduction process construct a machine that distinguishes whether another machine accepts a given input?” \\
\textit{Tabular queries} ask after data in rows, columns, or configuration tables like instruction sets or protocol specifications. For example, “What are the I/O port address ranges assigned to different devices in a typical PC configuration?” \\
\textit{Numerical queries} evaluate the retrieval of quantitative values such as thresholds, performance metrics, or complexity bounds. For instance, “What is the cost difference between the path found by the weighted A search and the path found by standard A in terms of percentage?” \\
\textit{Multi-page queries} ask after spatially separated content, such as an algorithm’s theory on one page and its pseudocode or diagram on the next. For example, “What are the key design trade-offs in distributed systems?” \\
\textit{Conceptual queries} focus on abstraction and higher-level reasoning, requiring the model to link both visual and textual domains for understanding complex ideas. An example is “What is preemptive scheduling and how does it differ from non-preemptive scheduling?” 

%\color{purple}Beth: how many queries in each category?  Overall discuss why 65 queries, and why this many in one category, this many in another, etc.)\color{black}

%\red{not clear how technical 5 from semantic-10 differ from those queries in baseline-75}\black{}
%navya’s added text
%\paragraph{\textbf{Semantic-10 Query Set}} The Semantic-10 benchmark has 10 queries, five \textit{technical} and five \textit{conceptual} questions. 

The benchmark includes factual and technical queries to test the system’s ability to retrieve content with high semantic specificity. Some will include a mix of visual or layout-based cues.  Other queries isolate domain-specific terminology and target precise definitions or algorithmic explanations without relying on structural hints (cues). The latter are well suited to assess whether the system can resolve fine-grained semantic meaning from embeddings alone, especially when the phrasing of the query does not directly match the surface text.

%Non-conceptual queries are structured to target well-defined and precise knowledge, such as “On which page is Chomsky Normal Form explained?”. Responses require high specificity and semantic matching. 
Conceptual queries, on the other hand, focus on abstraction and require the model to link higher-level reasoning across modalities. For instance, the two questions \textit{“Which page explains the intuition behind eventual consistency?”} and \textit{“Where is deadlock recovery illustrated with both text and diagram?”} test the model's ability to interpret questions that span both visual and textual domains.

\subsection{Query Set Validation} Each query is manually validated through a two-stage process. First, we manually verify that at least one relevant answer exists within the dataset of 3,612 textbook pages. Second, we refine query phrasing for clarity and specificity, ensuring that questions are not overly ambiguous or under-specified. Queries that fail these criteria are revised or excluded. This process ensures that all queries are clearly understandable and with verifiable answers in the corpus.

To maintain experimental consistency, every query is applied uniformly across all retrieval pipelines. The only changing variable across experiments, as mentioned earlier, is the similarity function used during vector comparison in Qdrant: Cosine, Dot Product, Euclidean, or Manhattan. All other components, including the embedding model (PaliGemma-3B), retrieval configuration, and evaluation metrics, are held constant. 
%This control allowed us to isolate and study the influence of each distance metric on the quality and contextual relevance of retrievals.

This structured and controlled approach to query design additionally allows for detailed performance analysis across semantic, structural, and visual dimensions. By anchoring the evaluation in both benchmark and semantic query sets, we ensure robustness, repeatability, and insight into metric-specific retrieval behavior across diverse query types.

\subsection{Evaluation Metrics}
%To evaluate the effectiveness of different similarity metrics on semantic document retrieval, we designed a comprehensive evaluation framework combining both quantitative metrics and qualitative analysis. Our goal was to understand not just which metric performs best numerically, but also which one retrieves the most contextually relevant content for educational use cases. All experiments were conducted on a fixed dataset of 3,612 page images using a consistent retrieval pipeline, with only the similarity metric varying across runs.

%We first compare the different distance measures. The similarity measure comparison is carried out on the Baseline-75 benchmark. Each query has a predetermined gold standard result that is manually curated. The purpose of the evaluation is to determine the best similarity measure on the Baseline-75 query benchmark.  

%We evaluate retrieval effectiveness using standard information retrieval metrics: \begin{itemize}
%    \item \textbf{Precision@3}: Proportion of relevant documents among the top-3 retrieved.
 %   \item \textbf{Recall@3}: Proportion of relevant documents successfully retrieved in the top-3.
  %  \item \textbf{F1 Score@3}: Harmonic mean of precision and recall.
   % \item \textbf{Average Precision (AP)}: Measures both relevance and ranking.
    %\item \textbf{Reciprocal Rank (RR)}: Inverse of the position of the first relevant result.
%\end{itemize}

We use a set of \textit{quantitative metrics} commonly employed in information retrieval tasks:

\begin{itemize}
    \item \textbf{Precision@5:} The fraction of the $top-5$ retrieved pages that are relevant (with respect to the ground truth).
    \item \textbf{Recall@5:} The proportion of all known relevant pages that appear in the $top-5$ results.
    \item \textbf{F1 Score@5:} The harmonic mean of Precision@5 and Recall@5, which provides a balanced view of both relevance and completeness.
    \item \textbf{Average Precision (AP):} A measure that accounts for both the relevance and ranking order of retrieved pages, assigning higher weight to relevant pages retrieved earlier.
    \item \textbf{Mean Reciprocal Rank (MRR):} The reciprocal of the rank at which the first relevant page is retrieved, averaged across all queries. This rewards systems that return highly relevant content earlier in the result list.
\end{itemize}

We adopt a $Top-5$ retrieval evaluation framework to reflect both real-world usage patterns and experimental constraints. Users interacting with educational retrieval systems typically focus on the first few results, expecting high-precision responses without needing to sift through large result sets. Thus, evaluating only the top 5 retrieved pages provides a realistic and user-centered measure of performance. Methodologically, it ensures consistent comparison across similarity functions—such as cosine and dot product—by minimizing the impact of their differing score distributions. The limited result set also makes it feasible to conduct detailed manual validation across the benchmark.

\subsection{Results}

\paragraph{Experiment I: Full Benchmark}
In this first experiment, we evaluate similarity functions using the full benchmark. The retrieval process utilizes \textit{late interaction} which processes queries and documents separately until the final stages of the retrieval process. This approach, popularized by the ColBERT ranking model, allows for more efficient and precise retrieval~\cite{khattab2020colbert}. 

Results are determined by summing up the resulting scores for all query tokens to yield a final relevance score per page. The $Top-5$ pages with the highest scores are selected as the \textit{retrieved results}. This Top-k selection is performed independently for each metric and query. The use of a fixed $k$-value ensures that evaluation metrics remain comparable across different scoring schemes, regardless of the absolute magnitude or range of scores produced by each similarity function.

The average metric scores across the full benchmark can be found in Table~\ref{tab:metric_comparison}.
%\color{purple}We want two versions of this table, one for  baseline-65 query set and one that isolates just the multi-page queries. Take the concept-10 queries out of these results.  \color{black}
While Dot Product comparably performs against Cosine in raw metrics, deeper inspection reveals its tendency to retrieve results biased by vector magnitude, occasionally surfacing less relevant but numerically high-scoring pages. In contrast, Cosine similarity offers a more consistent semantic signal. Cosine similarity is particularly effective for 1) abstract or paraphrased queries, 2) queries involving diagrams or tabular layouts, and 3) visually dense pages with minimal text.

\begin{table}[H]
\centering
\caption{Benchmark Comparison Across Similarity Functions}
\label{tab:metric_comparison}
\begin{tabular}{|l|c|c|c|c|}
\hline
\textbf{Metric}   & \textbf{Cosine} & \textbf{Dot} & \textbf{Euclidean} & \textbf{Manh} \\ \hline
Prec@5            & 0.514        & 0.496      & 0.234           & 0.418      \\ \hline
Recall@5          & 0.281        & 0.272      & 0.125          & 0.218      \\ \hline
F1@5              & 0.353       & 0.341      & 0.160          & 0.284      \\ \hline
AP          & 0.238        & 0.233     & 0.098       & 0.184      \\ \hline
MRR        & 0.801      & 0.787      & 0.418           & 0.738      \\ \hline
\end{tabular}
\end{table}

Euclidean and Manhattan distances perform significantly worse. Their sensitivity to both vector magnitude and dimensionality make them poorly suited for high-dimensional embeddings, often returning semantically incoherent results or textually similar but irrelevant pages.

We dive in on one specific query:
\textit{“How do state transition graphs work?”} from our use case queries in Sec~\ref{sec:usecase}. This is an example of a multi-modal query because state transition graphs are frequently presented both diagrammatically and described in narrative form. 
%While the gold standard included one page with the state graph diagram, our system retrieved a secondary page containing a paragraph-length textual explanation of each state transition. Though it lacked the diagram itself, the conceptual alignment between the text and the query intent made it highly relevant. \color{red}can we modify last sentence to this\color{black}
Within the Top-5 results is contained both the diagram and a textual explanation, with the diagram illustrating the state transition graph and the text providing a detailed explanation of each state transition. This combination of visual and textual content ensured a comprehensive response, fully aligning with the query's multi-modal intent.

No distance measure achieved strong precision accuracy, indicating that future work could explore page-level re-ranking models or context-aware retrieval strategies.

\paragraph{Experiment II: Multi-page Queries}

In this second experiment we use a subset of the full benchmark, the 14 multi-page queries, to see if the results change our conclusion with respect to the best distance measure.

Multi-page queries target a response that likely spans more than one page, for instance, a software algorithm that spans multiple pages, or a graph on one page and related text on another page.

\begin{table}[H]
\centering
\caption{Multipage Query Subset Comparison}
\label{tab:metric_comparison_multi-page}
\begin{tabular}{|l|c|c|c|c|}
\hline
\textbf{Metric}   & \textbf{Cosine} & \textbf{Dot} & \textbf{Euclidean} & \textbf{Manh} \\ \hline
Prec@5            & 0.457        & 0.442      & 0.157          & 0.342    \\ \hline
Recall@5          & 0.239        & 0.231      & 0.069           & 0.176      \\ \hline
F1@5              & 0.306       & 0.296      & 0.093           & 0.232      \\ \hline
AP          & 0.176        & 0.169      & 0.037           & 0.138      \\ \hline
MRR        & 0.690       & 0.654      & 0.246          & 0.571      \\ \hline
\end{tabular}
\end{table}

Results from our second experiment, see Table~\ref{tab:metric_comparison_multi-page}, shows the average metric scores across the multi-page queries. One can observe that cosine outperforms the other distance measures across the metrics.  Note the lower mean reciprocal rank (MRR) score compared to the full benchmark which holds across all of the distance measures. MRR only considers the rank of the first relevant answer and ignores possible further relevant answers so a lower MRR suggest that MRR is less meaningful for multi-page queries. 

%One can conclude a several things from the results. First, there is higher semantic continuity across retrieved pages, second, there is better relative ordering in multi-page contexts, and third, there are fewer out-of-context or redundant results. 

The query from our use cases in Sec~\ref{sec:usecase} \textit{“Tell me about Bellman-Ford algorithm and give an example?”} requires both the theoretical explanation and a worked example. As algorithms can take up considerable space on a page, it is reasonable to expect that the results are presented across two or more consecutive pages. Hence this query is an example of a multi-page query. 

Analyzing the results from this specific query, our ground truth labeled only the initial page with the algorithm description. However, our system retrieved a second page (not labeled in the ground truth) that included a diagram illustrating the step-by-step relaxation process on a sample graph. Human and GPT-4 validation confirmed its relevance. 

\paragraph{Experiment III: Conceptual Query Subset}
In our third experiment we use the 13 queries from the Conceptual subset of the benchmark.  As with Table~\ref{tab:metric_comparison_multi-page}, MRR is slightly better but still low compared to the full benchmark. Precision of the Top 5 and Average Precision are slightly better than for the multi-page subset but still below the overall benchmark. 

%\color{red}insert table here\color{black}
\begin{table}[H]
\centering
\caption{Conceptual Query Subset Comparison}
\label{tab:metric_comparison_conceptual}
\begin{tabular}{|l|c|c|c|c|}
\hline
\textbf{Metric}   & \textbf{Cosine} & \textbf{Dot} & \textbf{Euclidean} & \textbf{Manh} \\ \hline
Prec@5            & 0.461        & 0.400      & 0.184          & 0.323      \\ \hline
Recall@5          & 0.226        & 0.198      & 0.087           & 0.157     \\ \hline
F1@5              & 0.303        & 0.264      & 0.118           & 0.210      \\ \hline
AP          & 0.182        & 0.164      & 0.059           & 0.126      \\ \hline
MRR        & 0.692       & 0.653      & 0.358           & 0.673      \\ \hline
\end{tabular}
\end{table}

\paragraph{Experiment IV: Unbounded Analysis}
Our final experiment studies relevant but unlabeled retrievals. This is where we look beyond the Top-5 results. While the Top-5 results provides us with a focused and comparable assessment, it necessarily leaves out information about the tail of the similarity distribution. To address this limitation, we carry out an unbounded retrieval analysis, where we retrieve all indexed documents and plot similarity scores across the full ranked list.

This experiment again uses the full benchmark to explore the retrieval system’s behavior in response to challenging queries—especially those that may surface relevant content overlooked by the ground truth. By expanding our evaluation beyond Top-5 and beyond rigid labels, we aim to learn how well the system generalizes in less constrained settings.

\begin{figure}[htbp]    % Changed to figure* for full width
    \centering
    \includegraphics[width=7cm]{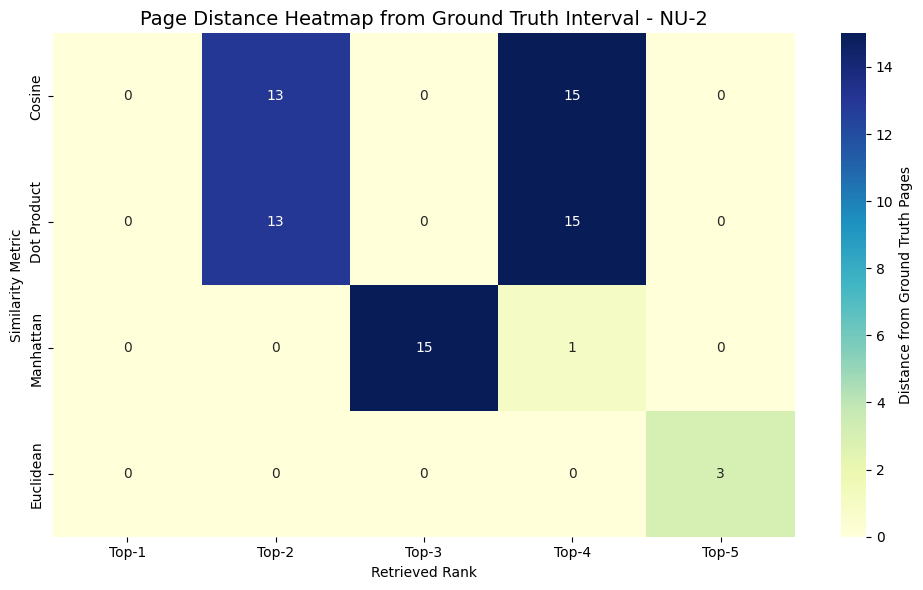}
    \caption{Page distance for numerical query NU-2}
    \label{fig:heatmap}
\end{figure}

%\color{blue}describe figures (have more than one).\color{black}

We conduct this evaluation using both Cosine similarity and Dot Product which are the top two performing distance functions from our earlier experiments. For each query, we identify pages returned outside the Top-5 that exhibit strong semantic alignment with the query but are not present in the original ground truth set. %We classify these as \textit{relevant-but-unlabeled}. 
To assess their value, we:

\begin{itemize}
    \item 1) Measure similarity scores between the query and the retrieved page embeddings using both Cosine and Dot Product.
    \item 2) Perform manual annotation and GPT-4-based validation to confirm contextual alignment.
\end{itemize}

The second step exposes limitations in rigid evaluation and highlighted the strengths of angular-based metrics like Cosine Similarity, as well as the density-driven retrieval capacity of Dot Product. Both metrics retrieve high-scoring pages that are semantically aligned with the intent of the query, even when those pages were not labeled in the gold set.

In total, 18 out of 75 queries return relevant-but-unlabeled results across the two metrics. These pages had Cosine or Dot Product similarity scores greater than 0.85, and 72\% were validated as relevant by both human annotators and GPT-4. Many of these cases were associated with conceptual queries — those that required abstract reasoning, inference, or multimodal alignment.

We illustrate results in Figure ~\ref{fig:heatmap} using a single benchmark query.  The result is from query NU-2 (numerical query subset, query \#2):   \textit{What is the cost difference between the path found by the weighted A search and the path found by standard A in terms of percentage}.  %beth: not sure this heatmap supports the conclusion.  Narrative example map be better.

The query, presented first in Sec~\ref{sec:usecase}, is this: 
\textit{“Which page explains the intuition behind eventual consistency?”}. This query is not tied to keywords or diagrams but to deeper conceptual reasoning. The labeled ground truth included a section heading titled “Eventual Consistency.” However, our system surfaced an unlabeled page that drew analogies to distributed clocks and causal ordering, both of which are semantically aligned but not explicitly labeled. This page received high relevance scores from Cosine similarity and was validated as meaningful in explaining the intuition behind eventual consistency. %Such flexible interpretation is crucial in domains like philosophy, political theory, or computer science where conceptual richness extends across context.

This final experiment exposes a key insight: strict label-based evaluation may overlook true semantic matches, particularly in educational material with distributed or inferred knowledge. Evaluating retrieval effectiveness solely by exact matches in the Top-k list underrepresents a system's potential, especially when applied to open-ended, conceptual information needs. Both Cosine and Dot Product demonstrate success in generalizing beyond the ground truth, surfacing valuable information.

\section{Discussion}\label{sec:discussion}

%Navya, add 3 query examples here. as we want queries that are interesting to HathiTrust, we want 1) multi-page, 2) multi-modal, and 3) concept driven.  give the query. For results, we want one result (at a minimum) drawn from "relevant but unlabeled".   Give examples of responses that fall outside the manual gold standard. 

Digital library collections from research libraries contain high-density visual information (e.g., illustrations, maps, mathematical diagrams), multi-page narratives, and domain-specific abstractions that cannot be captured effectively through keyword search or full-text indexing alone. Our work strives to illustrate the strengths and weaknesses of vector embedding as a means to expand the toolset for content discovery.  

%of multi-modal retrieval systems powered by vision-language models offer a promising alternative by enabling semantic and visual discovery directly over page-level image embeddings.

Our results are to some extent inconclusive and suggest further investigation.  We have shown statistically that the strongest distance measure, cosine similarity, will retrieve results that fall within the gold standard with a precision of slightly over 0.514. This means that a little over half of the top-5 retrieved pages for the benchmark fall within the ground truth set. The F1 score, which provides a balanced view of relevance and completeness is weaker, coming in at .353 for the full benchmark. 

But these numbers are relative to the ground truth dataset which is a manually curated dataset that was assembled by graduate students over the course of a couple of months.  The next step in this work is strengthen the completeness and relevance of the ground truth itself. 

It is also interesting that the Mean Reciprocal Rank (MRR) is lower for both query subsets than it is for the benchmark as a whole.   This needs further investigation. 

 Our benchmark and study reflect a broader insight that is a strength of this work: conventional gold-standard labels, while necessary for benchmarking, may miss semantically rich results that lie outside literal keyword overlap or manually annotated sections. A retrieval system that generalizes across pages, modalities, and conceptual space can significantly enhance user experience in digital libraries. 

The other aspect of the study is architectural. Suppose we treat the vector database as a standalone resource for discovery. This is reasonable as Colpali supports late interaction, which means that one can do the vector embedding of all the texts at once, and at a later time and on demand come back and embed the queries. But for our studies, the queries had to be embedded one at a time and the results measured.  This means that every query has to be embedded by itself (no bulk embedding of queries) before it can return results.  The query response time is thus dependent on the level of concurrency supported within the multi-modal model.  That is, can the multi-modal model support multiple query embeddings at the same time?  This is an area of future study for us.

\section{Related work}\label{sec:related}
Document retrieval systems have evolved beyond simple text-based approaches to incorporate visual elements and leverage advanced language models. Traditional methods rely primarily on statistical techniques such as TF-IDF~\cite{sparckjones1972statistical} and BM25~\cite{robertson1994okapi}, which were efficient but limited in their ability to capture semantic meaning and visual context. With the rise of deep learning, transformer-based language models have enabled semantically richer embeddings, supporting more effective document retrieval over natural language queries.

Multimodal learning for document retrieval has gained traction with the introduction of vision-language models (VLMs) that integrate visual layout, image regions, and text sequences. Models like LayoutLM~\cite{huang2022layoutlmv3pretrainingdocumentai} and DocFormer~\cite{appalaraju2021docformerendtoendtransformerdocument} attempt to represent these modalities jointly through a single embedding vector. %However, such early interaction models often lose fine-grained alignment needed to capture token-level or patch-level relationships. 
%navya added text
However, because early interaction models like LayoutLM and DocFormer combine textual and visual information into a single embedding early in the processing pipeline, they often lose the ability to maintain fine-grained alignment—meaning the model cannot precisely distinguish or link individual tokens (such as words or symbols) to specific visual elements (such as regions of a diagram, table cells, or bounding boxes). This lack of detailed correspondence makes it difficult for the model to capture localized relationships between specific query terms and document content, which is essential for tasks like retrieving a diagram that corresponds to a particular labeled concept or identifying a figure that explains a formula mentioned in the query.
%Additionally, these models are typically benchmarked on scanned document collections such as RVL-CDIP and FUNSD whereas we are focused on textbooks.

ColPali~\cite{faysse2024colpali} is a recent multimodal retrieval model designed to address these limitations. It employs multi-vector representations and late interaction scoring to retrieve content from documents with dense visual and textual information. Its ability to handle tables, diagrams, and complex layouts with high indexing throughput makes it suitable for real-world applications, including search engines and Retrieval-Augmented Generation (RAG) systems~\cite{Collell_Zhang_Moens_2017}.

While late interaction frameworks such as ColBERT~\cite{khattab2020colbert} have demonstrated success in text-based retrieval, they are typically optimized for short-form content like web passages. Our study extends this retrieval paradigm to visually complex, full-page image inputs in digitized books. We also introduce novel benchmark sets specifically designed to test both semantic and cross-modal retrieval performance. This controlled evaluation—centered around similarity function analysis—addresses key gaps in existing systems, especially those intended for large-scale digital library collections.

\section{Conclusion}\label{sec:conclusion}
By holding the retrieval model, query set, and indexing method constant and varying only the similarity function, we isolated the contribution of the distance metric to retrieval effectiveness. This modular approach also facilitates clear attribution of performance trends, enabling insight into how different metrics influence semantic alignment and ranking behavior across varied query types.

%No metric achieved strong position accuracy, indicating that future work could explore page-level re-ranking models or context-aware retrieval strategies to improve sequence fidelity.

The results presented here support the case for augmenting traditional search with multimodal vector retrieval, particularly in large-scale collections where diagrammatic and abstract content is prevalent.

The Vector Benchmark result set can be found here \cite{resultsetPlale} and the code used in this study can be found here \cite{githubrepoPlale}.

\section{Acknowledgements}
With thanks to Manikanta Kodandapani Naidu for contributions to the software framework and Neelesh .

% \addbibresource{bibliography.bib}
% \printbibliography[heading=bibintoc, title={References}]

\bibliographystyle{plain}
\bibliography{vector_embedding.bib}

\end{document}